\documentclass[aps,pre,twocolumn]{revtex4-1}

\usepackage{amsmath}
\usepackage{amssymb}
\usepackage{color}
\usepackage{hyperref}

\usepackage{graphicx}

\DeclareMathAlphabet{\mathitbf}{OML}{cmm}{b}{it}

\newcommand{\ket}[1]{|#1\rangle}
\newcommand{\bra}[1]{\langle #1|}
\newcommand{\braket}[2]{\langle #1|#2\rangle}

\newcommand{\dbar}{{\,\mathchar'26\mkern-12mu d}}

\newcommand{\sFrac}[2]{{\textstyle\frac{#1}{#2}}}

\begin{document}
\title{Quasilocalized states of self stress in packing-derived networks}
\author{Edan Lerner}
\affiliation{Institute for Theoretical Physics, University of Amsterdam, Science Park 904, 1098 XH Amsterdam, The Netherlands}

\begin{abstract}
States of self stress (SSS) are assignments of forces on the edges of a network that satisfy mechanical equilibrium in the absence of external forces. In this work we show that a particular class of quasilocalized SSS in packing-derived networks, first introduced in [D.~M.~Sussman, C.~P.~Goodrich, and A. J. Liu, Soft Matter \textbf{12}, 3982 (2016)], are characterized by a lengthscale $\ell_c$ that scales as $1/\sqrt{z_c-z}$ where $z$ is the mean connectivity of the network, and $z_c\!\equiv\!4$ is the Maxwell threshold in two dimensions, at odds with previous claims. Our results verify the previously proposed analogy between quasilocalized SSS and the mechanical response to a local dipolar force in random networks of relaxed Hookean springs. We show that the normalization factor that distinguishes between quasilocalized SSS and the response to a local dipole constitutes a measure of the mechanical coupling of the forced spring to the elastic network in which it is embedded. We further demonstrate that the lengthscale that characterizes quasilocalized SSS does not depend on its associated degree of mechanical coupling, but instead only on the network connectivity.

\end{abstract}

\maketitle

\section{introduction}
The unjamming point \cite{ohern2003,van_hecke,liu_review,matthieu_thesis} marks the loss of solidity in disordered materials that occurs by tuning some external, macroscopic (e.g.~deformation or confining pressure) \cite{ohern2003} or intrinsic, microscopic (e.g.~the connectivity of a network) control parameter \cite{wouter_epl_2009}. While substantial progress in understanding the nature of the unjamming transition has been achieved in recent years \cite{eric_emt_prestress, eric_finite_T_jamming, parisi_fractal, Zamponi_2014_full_thing,berthier_hard_spheres_swap}, several aspects of this critical point are still debated. 

One of the enduring open problems within the field of unjamming concerns the identification of the various diverging lengthscales associated with this transition, and understanding their dependencies on the relevant control parameters. Most previous observations focus on two lengthscales, which follow different scaling laws with connectivity $z$; the first length $l^*\!\sim\!(z\!-\! z_c)^{-1}$ with $z_c\!\equiv\!2\dbar$ in $\dbar$ spatial dimensions was first put forward in \cite{wyart_witten_epl_2005,matthieu_thesis}. It emerges due to an interplay between boundary constraints and bulk degrees of freedom. A point-to-set correlation length that follows the same scaling $\!\sim\!(z\!-\! z_c)^{-1}$ was extracted in \cite{bulbul_2011_ell_star} by fixing the forces that cross the boundary of a square cavity in a packing and analyzing the force-balance solutions inside the cavity. The length $l^*$ was further identified in floppy materials \cite{phonon_gap} by freezing the degrees of freedom outside a spherical shell, and decreasing the size of the shell until the floppiness of the interior of the shell disappears. A dual protocol was carried out in soft sphere packings in two dimensions, then the interactions across the boundaries of a square zone were eliminated, and the size of the zone was reduced until rigidity within the zone was lost \cite{goodrich_ell_star_soft_matter_2013}. In \cite{ellenbroek_point_response_2009} it was claimed that the length $l^*$ can be observed by considering the mechanical response to inflating a single particle in a packing of soft spheres. In \cite{maloney_2015_lengthscales_PRE} $l^*$ was argued to control fluctuations in coarse-grained elastic moduli fields. More recently, the fluctuations of the mechanical response to nonlocal forcing in soft-sphere packings were analyzed and shown to exhibit a signature of $l^*$ \cite{brian_length_prl_2017}.

The second length $\ell_c\!\sim\!(z\!-\! z_c)^{-1/2}$ associated with the unjamming transition characterizes the mechanical response to various local and global perturbations, and was shown to mark the crossover between atomistic-scale and continuum-like mechanical responses. The length $\ell_c$ was first observed in \cite{silbert2005} by extracting the dominant wavelength of vibrational modes of a packing of soft spheres at the frequency scale $\omega^*\!\sim\! z\!-\! z_c$, in consistency with later theoretical predictions \cite{mw_EM_epl} using effective medium theory. In \cite{phonon_gap} the length $\ell_c$ was predicted to characterize the response to local perturbations in floppy spring networks using similar theoretical tools. In \cite{biroli_jamming_jcp_2013} the length $\ell_c$ was observed by considering a rescaled Debye-Waller factor in harmonic spheres at vanishing temperatures above and below the jamming point. In \cite{goodrich_ell_c_soft_matter_2013} $\ell_c$ was observed in the linear response to boundary perturbation in two and three dimensions. A more direct observation of $\ell_c$ was made in \cite{breakdown} by considering the mechanical response to local dipole forces in packings of harmonic discs and random networks of Hookean springs. In \cite{maloney_2015_lengthscales_PRE} $\ell_c$ was shown to characterize the transverse response to a point force. In \cite{silbert_prl_2016_length} and \cite{Massimo_length_2016} $\ell_c$ was identified by identifying the crossover to continuum like fluctuations of coarse-grained elastic moduli. More recently, the transverse nonaffine displacements fluctuations in response to long-wavelength forcing were shown to exhibit the scale $\ell_c$ in \cite{brian_length_prl_2017}. 

Other diverging lengths besides those mentioned above have been identified in previous studies of jamming and unjamming; some examples are the correlation length of non-affine displacements observed in strain-stiffening floppy networks, shown in \cite{asm_self_organization_strain_stiffening} to scale as $(\gamma_c\!-\!\gamma)^{-1/2}$ in networks deformed away from the stiffening strain $\gamma_c$. In \cite{maloney_2015_lengthscales_PRE} a length $\sim\!(z\!-\! z_c)^{-0.4}$ was observed in the longitudinal response to a point forcing in harmonic disc packings. In \cite{brian_length_prl_2017} a length $\sim\!(z\!-\! z_c)^{-0.66}$ was shown to describe the longitudinal compliance in harmonic sphere packings. In \cite{andreotti_prl_nonlocal_2013} a length that characterizes nonlocality in granular flows was observed to grow with decreasing stress anisotropy $\mu$ towards the critical $\mu_c$ as $(\mu\!-\!\mu_c)^{-1/2}$.

In this work we study the lengthscale that characterizes quasilocalized states of self stress  as observed in packing-derived contact networks in two dimensions. States of self stress (SSS) are assignments of contractile or compressive forces on the edges of a network, that satisfy mechanical equilibrium on the nodes of the network \cite{calladine1978}. They play an important role in determining the force chains in granular matter \cite{PhysRevLett.116.078001} and the physics of topological metamaterials \cite{Kane2014,Paulose2015}. In random networks with connectivities above the isostatic point $z_c$ one expects the number of orthonormal SSS to be extensive \cite{sussman_sss}, and proportional to $z\!-\! z_c$, if fluctuations in the connectivity of the network are small \cite{ellenbroek_rigidity_prl_2015}. 

In \cite{sussman_sss} (referred to as SUS in what follows) a particular construction of orthonormal set of SSS was introduced (see precise definitions in what follows); the set can be defined given a choice of any edge of the network, such that one SSS is quasilocalized on that particular edge, and all other SSS have no component on that edge. In what follows we refer to the quasilocalized member of the constructed orthonormal set of SSS as a \emph{quasilocalized state of self stress} (QLS). The spatial decay of QLS was argued in SUS to be characterized by a length $\ell_{\mbox{\tiny SSS}}\!\sim\!(z\!-\! z_c)^{-0.8}$ in two dimensions (2D), and $\ell_{\mbox{\tiny SSS}}\!\sim\!(z\!-\! z_c)^{-0.6}$ in three dimensions (3D). In \cite{comment_liu} an explicit expression for QLS was put forward, from which a relation between QLS in random networks and the mechanical response to local dipolar forces (referred to in what follows simply as the \emph{dipole response}) in Hookean spring networks can be directly established. Based on this relation, it was argued in  \cite{comment_liu} that the spatial decay of SSS should be characterized by the same length $\ell_c\!\sim\!(z\!-\! z_c)^{-1/2}$ that characterizes dipole responses as observed in \cite{breakdown}, and at odds with the observations of \cite{sussman_sss}. In \cite{reply_to_comment} this disagreement is discussed, and the authors conclude that it is ``open to interpretation" which of the scaling laws $(z\!-\! z_c)^{-1/2}$ or $(z\!-\! z_c)^{-2/3}$ better describes their data for the spatial decay of QLS. It is further suggested in \cite{reply_to_comment} that the normalization factors (defined and discussed in detail in what follows) that distinguish between QLS and dipole responses, which lead to a different ensemble averaging of these two objects, could alter the scaling with connectivity of the lengthscale that characterizes these objects' spatial decay. 

Here we resolve the controversy described above and provide direct numerical evidence that the lengthscale that characterizes the spatial decay of QLS in packing-derived networks is indeed $\ell_c\!\sim\!(z\!-\! z_c)^{-1/2}$, as proposed in \cite{comment_liu}, and at odds with the observations of \cite{sussman_sss} and with the claims made in \cite{reply_to_comment}. We go further and confirm that the normalization factors that were neglected in the argumentation of \cite{comment_liu} do indeed not affect the scaling with connectivity of the discussed lengthscales. We show that the spatial decay of QLS with very large normalization factors exhibits the same lengthscale as the spatial decay of QLS with typical normalization factors, and further demonstrate the lack of correlation between this lengthscale and normalization factors by examining the spatial patterns of dipole responses.

Our work is structured as follows; in Sect.~\ref{methods_section} we provide details of the models studied and numerical methods used. In Sect.~\ref{theory} we review the theoretical formalism presented in \cite{comment_liu} and \cite{breakdown}, within which QLS and the mechanical responses to local dipolar forces in relaxed Hookean spring networks are defined, and we discuss various mechanical interpretations of the normalization factors associated to QLS. In Sect.~\ref{results} we show data from our numerical simulations that supports that the lengthscale that characterizes the spatial decay of QLS is $\ell_c\!\sim\!(z\!-\! z_c)^{-1/2}$. Our findings are summarized in Sect.~\ref{discussion}.

\section{Models and methods}
\label{methods_section}

In this work we study random networks in 2D derived from the underlying network of contacts between soft discs in large two-dimensional packings. Our soft discs interact via the pairwise potential
\begin{equation}
\varepsilon(r_{ij}) = \Theta\big((\rho_i + \rho_j) - r_{ij}\big)\sFrac{\kappa}{2}\big(r_{ij} - (\rho_i + \rho_j)\big)^2\,,
\end{equation}
where $\rho_i$ denotes the radius of the $i^{\mbox{\tiny th}}$ particle, $\kappa$ is a stiffness (set to unity in what follows), $r_{ij}$ is the pairwise distance between the centers of particles $i$ and $j$, and $\Theta(x)$ is the heaviside step function. All particles share the same mass $m$ (also set to unity). We created packings of up to $N\!=\!10^6$ particles, half of which have a radius of $\rho\!=\! 0.5$ and the other half have $\rho\!=\! 0.7$. Distances are measured in terms of the diameter $D$ of the smaller particles, energies in terms of $\kappa D^2$, and stresses in terms of $\kappa$. The key control parameter for our packings is the pressure, which is set by applying compressive or decompressive strain in small steps, followed by a relaxation of the potential energy by means of the FIRE algorithm \cite{fire}. We created packings at pressures ranging from $p\!=\!10^{-1}$ to $p\!=\!10^{-5}$, where the highest pressure states were created by relaxing a random configuration, and subsequent lower pressure packings were created by manipulating the higher pressure packings. A packing is deemed relaxed once the ratio of the typical net force on the particles to the pressure drops below $10^{-7}$. The connectivity $z$ is measured in each packing by eliminating `rattler' particles from the analysis as described in \cite{ASM_simulations_paper}. In what follows we solve linear systems of equations by a conventional conjugate gradient solver.

\section{Quasilocalized states of self stress and dipole responses}
\label{theory}

In this section we review the theoretical framework \cite{breakdown,comment_liu,calladine1978,matthieu_thesis} within which the two objects of interest -- dipole responses and QLS -- are defined and can be related. We also hold a discussion about the normalization factors that are shown to distinguish between dipole responses and QLS.

\subsection{Response to a local dipolar force in Hookean spring networks}

We consider a random network of unit point masses connected by relaxed Hookean springs, i.e.~that all springs resides precisely at their respective rest-lengths, so that the energy of the mechanical equilibrium ground state is zero. We assume that the network connectivity $z\!>\! z_c$, and that all the springs share the same stiffness $\kappa$, which together with the characteristic length $\lambda$ of a spring forms our microscopic unit of energy $\kappa\lambda^2$. We label springs by Greek letters, and coordinates by Roman letters. The potential energy reads
\begin{equation}
U = \sFrac{1}{2}\sum_\alpha (r_\alpha - \ell_\alpha)^2\,,
\end{equation}
where we have set $\kappa\!=\!1$, $r_\alpha$ is the length of the $\alpha^{\mbox{\tiny th}}$ spring, and $\ell_\alpha$ is its rest-length. The dynamical matrix reads
\begin{equation}\label{foo00}
{\cal M} \equiv \frac{\partial^2U}{\partial \vec{x}\partial \vec{x}} = \sum_\alpha \vec{D}_\alpha\vec{D}_\alpha\,,
\end{equation}
where we have introduced the dipole vectors $\vec{D}_\alpha\!\equiv\!\frac{\partial r_\alpha}{\partial\vec{x}}$. Notice that since we consider relaxed spring networks, the term that involves tensions or compressions in the springs is absent from Eq.~(\ref{foo00}). The dynamical matrix can be expressed in terms of the equilibrium matrix ${\cal S}$ \cite{calladine1978} as
\begin{equation}\label{foo02}
{\cal M} = {\cal S}^T{\cal S}\,.
\end{equation}
The equilibrium matrix ${\cal S}$ holds geometric information of the spring network. It is related to the dipole vectors $\vec{D}_\alpha$ via
\begin{equation}
\vec{D}_\alpha \leftrightarrow {\cal S}^T\ket{\alpha}\,,
\end{equation}
where $\ket{\alpha}$ is a vector in the space of springs which has zeros in all components except for the $\alpha^{\mbox{\tiny th}}$ component which is set to unity. If a dipolar force $\vec{D}_\alpha$ is applied to the network, the (linear) displacement response reads 
\begin{equation}\label{foo15}
\delta \vec{R} = {\cal M}^{-1}\cdot\vec{D}_\alpha\,,
\end{equation}
written in bra-ket notation as
\begin{equation}\label{foo11}
\ket{\delta R} = {\cal M}^{-1}{\cal S}^T\ket{\alpha}\,.
\end{equation}
We denote by $\ket{\varphi}$ the set of forces that arise in the springs due to the displacement $\ket{\delta R}$, referred to in what follows as the dipole response. In our system of Hookean springs with unit stiffnesses, and to linear order in $\ket{\delta R}$, these are simply the elongation or contraction of each spring, namely
\begin{equation}\label{foo10}
\ket{\varphi} = {\cal S}\ket{\delta R} = {\cal S}\big({\cal S}^T{\cal S}\big)^{-1}{\cal S}^T\ket{\alpha}\,,
\end{equation}
where we have used Eqs.~(\ref{foo02}) and (\ref{foo11}). This expression for the dipole response $\ket{\varphi}$ will be compared to analogous expressions for QLS in what follows.

\subsection{Quasilocalized states of self stress}

We consider next networks where each edge is thought of as a rigid bar, and we assume the connectivity is larger than the Maxwell threshold $z_c$. Such networks are referred to by some workers (e.g.~\cite{calladine1978}) as frames. States of self stress (SSS) are assignments $\{\phi_{jk}\}$ of forces on each of the edges $\langle jk\rangle$ of such a network, that satisfy mechanical equilibrium, namely that
\begin{equation}\label{foo04}
\vec{F}_k = \sum_{j(k)}\hat{n}_{jk}\phi_{jk} = 0\,,
\end{equation}
where $j(k)$ denotes all the nodes $j$ connected to the $k^{\mbox{\tiny th}}$ node, and $\hat{n}_{jk}$ is the unit vector pointing from the $j^{\mbox{\tiny th}}$ to the $k^{\mbox{\tiny th}}$ node. It is convenient to express Eq.~(\ref{foo04}) using our bra-ket notation as
\begin{equation}
\ket{F} = {\cal S}^T\ket{\phi}\,,
\end{equation}
where ${\cal S}$ is the same equilibrium matrix discussed above. If fluctuations in the connectivity $z$ of the network are small (see relevant discussion in \cite{ellenbroek_rigidity_prl_2015}), as assumed here and in what follows, the dimension of the null-space of ${\cal S}$ scales as $N(z\!-\! z_c)$ where here $N$ is the number of nodes in the network. In other words, zero is an eigenvalue of the operator ${\cal S}{\cal S}^T$, and there are on the order of $N(z\!-\! z_c)$ degenerate eigenmodes $\ket{\phi_\ell}$ of ${\cal S}{\cal S}^T$ associated with the zero eigenvalue, which precisely constitute a set of orthonormal SSS. We refer any such orthonormal set of solutions to Eq.~(\ref{foo04}) as a spanning of the null space of ${\cal S}{\cal S}^T$, or just a \emph{spanning set}. 

In SUS a particular spanning set was introduced as follows: given a choice of a single edge $\alpha$ of the network, all besides one member of the spanning set have no projection on the $\alpha^{\mbox{\tiny th}}$ edge. It was shown in SUS that the single member in this spanning set that has a nonzero projection on the $\alpha^{\mbox{\tiny th}}$ edge is quasilocalized, i.e.~its spatial structure is characterized by a core of size $\ell_{\mbox{\tiny SSS}}$, decorated by power-law decays in the far field. We therefore refer to such SSS as \emph{quasilocalized states of self stress} (QLS). A different and unique QLS can be associated with each edge $\alpha$ of the network. 

In \cite{comment_liu} it was shown precisely how to construct the QLS associated to any given edge $\alpha$ of a network. Here we briefly repeat that construction for completeness. We consider the network that remains after removing the $\alpha^{\mbox{\tiny th}}$ edge, and decorate with a tilde $(\sim)$ quantities defined on the network after removal of the $\alpha^{\mbox{\tiny th}}$ edge. We next define the set of edge forces $\ket{\tilde{f}^{(\alpha)}}$ that balance a dipolar force $\vec{D}_\alpha \equiv \frac{\partial r_\alpha}{\partial\vec{x}}$ (with $r_\alpha$ the length of the removed edge and $\vec{x}$ the nodes' coordinates) applied on the nodes that were connected by the $\alpha^{\mbox{\tiny th}}$ edge before its removal, namely
\begin{equation}\label{foo06}
\tilde{\cal S}^T\ket{\tilde{f}^{(\alpha)}} = -\ket{D_\alpha}\,.
\end{equation}
Operating on this equation with $\tilde{\cal S}$ and inverting it in favor of $\ket{\tilde{f}^{(\alpha)}}$ we obtain
\begin{equation}\label{foo05}
\ket{\tilde{f}^{(\alpha)}} = -\big(\tilde{\cal S}\tilde{\cal S}^T\big)^{-1}\tilde{\cal S}\ket{D_\alpha}\,,
\end{equation}
where the superscript $\circ^{-1}$ here and in what follows should be understood as the pseudo-inverse of a matrix wherever applicable. We note that Eq.~(\ref{foo05}) uniquely defines the assignment of forces $\ket{\tilde{f}^{(\alpha)}}$ on the edges of the network that balance the dipolar force $\vec{D}_\alpha$. 

In order to construct the spanning as introduced in SUS, we reconnect the removed edge $\alpha$ at its original location, and first construct the QLS $\ket{\phi_q}$ as follows: we calculate a normalization factor
\begin{equation}\label{foo14}
g_\alpha\equiv\big(\braket{\tilde{f}^{(\alpha)}}{\tilde{f}^{(\alpha)}}+1\big)^{-1/2}\,,
\end{equation}
and assign for every edge $\beta\!\ne\!\alpha$, $\braket{\phi_q}{\beta}\!=\! g_\alpha\braket{\tilde{f}^{(\alpha)}}{\beta}$, and finally we set $\braket{\phi_q}{\beta}\!=\! g_\alpha$. The rest of the members of the spanning set are obtained by considering any spanning set $\{\ket{\tilde{\phi}_\ell}\}$ of $\tilde{\cal S}\tilde{\cal S}^T$ and assigning zero to the additional $\alpha^{\mbox{\tiny th}}$ component of each member. 

It is immediately verified that the construction described above is precisely the construction introduced by SUS; for any $\ell\!\ne\! q$, $\braket{\phi_\ell}{\alpha}\!=\!0$, and $\braket{\phi_q}{\alpha}\!\ne\!0$, both by construction. Furthermore, Eq.~(\ref{foo05}) implies that $\ket{\tilde{f}^{(\alpha)}}$ is a superposition of nonzero modes of $\tilde{\cal S}\tilde{\cal S}^T$, and therefore $\braket{\tilde{f}^{(\alpha)}}{\tilde{\phi}_\ell}\!=\!0$, then for $\ell\!\ne\! q$
\begin{equation}
\braket{\phi_q}{\phi_\ell} = g_\alpha\big(\braket{\tilde{f}^{(\alpha)}}{\tilde{\phi}_\ell} + \braket{\alpha}{\phi_\ell}\big) = 0\,,
\end{equation}
as required. Finally, following Eq.~(\ref{foo06}) 
\begin{eqnarray}
{\cal S}^T\ket{\phi_q} & = & g_\alpha\big(\tilde{\cal S}^T\ket{\tilde{f}^{(\alpha)}} + {\cal S}^T\ket{\alpha} \big) \nonumber\\
& = & g_\alpha\big(\tilde{\cal S}^T\ket{\tilde{f}^{(\alpha)}} + \ket{D_\alpha}\big) = 0\,. 
\end{eqnarray}

Another definition of the QLS $\ket{\phi_q}$ is obtained by using our constructed set of SSS as described above, and writing
\begin{equation}\label{foo08}
\ket{\phi_q} \propto \sum_\ell \braket{\phi_\ell}{\alpha}\ket{\phi_\ell} = \bigg( \sum_\ell \ket{\phi_\ell}\bra{\phi_\ell}\bigg)\ket{\alpha}\,,
\end{equation}
where the sum runs over all the SSS, i.e.~the zero modes of ${\cal S}{\cal S}^T$, and notice that the above is merely a proportionality relation and not an equation. We next denote $\sum_m \ket{\phi_m}\bra{\phi_m}$ as the sum over outer products of \emph{nonzero} modes of ${\cal S}{\cal S}^T\!$; with this definition, one has
\begin{equation}\label{foo07}
\sum_\ell \ket{\phi_\ell}\bra{\phi_\ell} = {\cal I} - \sum_m\ket{\phi_m}\bra{\phi_m}\,,
\end{equation}
which is a projection operator onto the space that is orthogonal to the null-space of ${\cal S}^T$. In order to relate it the equilibrium matrix ${\cal S}$ itself, notice that if $z\!>\! z_c$ the nonzero modes $\ket{\phi_m}$ of ${\cal S}{\cal S}^T\!$ are related to the eigenmodes $\ket{\Psi_m}$ of ${\cal S}^T\!{\cal S}$ via \cite{asm_pnas}
\begin{equation}
{\cal S}\ket{\Psi_m} = \omega_m\ket{\phi_m}\,,
\end{equation}
where $\omega_m^2$ is the eigenvalue associated to $\ket{\Psi_m}$, and therefore
\begin{equation}
\sum_m\!\ket{\phi_m}\bra{\phi_m} = {\cal S}\!\left(\!\sum_m\!\frac{\ket{\Psi_m}\bra{\Psi_m}}{\omega_m^2}\!\right)\!{\cal S}^T\! = {\cal S}\big({\cal S}^T\!{\cal S}\big)^{-1}\!{\cal S}^T\,.
\end{equation}
Using this relation together with Eq.~(\ref{foo07}), we obtain
\begin{equation}
\sum_\ell \ket{\phi_\ell}\bra{\phi_\ell} = {\cal I} - {\cal S}\big({\cal S}^T\!{\cal S}\big)^{-1}{\cal S}^T\,.
\end{equation}
An expression for QLS follows from Eq.~(\ref{foo08}) as
\begin{equation}\label{foo09}
\ket{\phi_q} = \frac{\big({\cal I} - {\cal S}\big({\cal S}^T\!{\cal S}\big)^{-1}{\cal S}^T\big)\ket{\alpha}}{\sqrt{\bra{\alpha}{\cal I} - {\cal S}\big({\cal S}^T\!{\cal S}\big)^{-1}{\cal S}^T\ket{\alpha}}}\,.
\end{equation}

Eq.~(\ref{foo09}) constitutes a second, explicit definition of QLS, which is entirely equivalent to the construction based on Eqs.~(\ref{foo05}) and (\ref{foo14}). We have verified numerically that the two definitions exactly agree. Finally, by comparing Eqs.~(\ref{foo10}) and (\ref{foo09}), it is clear that for edges $\beta\!\ne\!\alpha$, $\braket{\beta}{\phi_q}\!\propto\!\braket{\beta}{\varphi}$, i.e.~the dipole response $\ket{\varphi}$ is proportional to the QLS $\ket{\phi_q}$, except for their $\alpha^{\mbox{\tiny th}}$ components. The proportionality constant that separates the two objects is the normalization factor, denoted by $c_\alpha$ and discussed in detail below.

\subsection{Normalization factors of QLS}

In \cite{reply_to_comment} Eq.~(\ref{foo08}) was suggested as the definition of $\ket{\phi_q}$, together with a declaration that normalization factors were neglected for the sake of brevity. Notice that the relevant normalization factors $c_\alpha$ are different than the normalization factors $g_\alpha$ defined by Eq.~(\ref{foo14}). Instead, they read
\begin{equation}\label{foo12}
c_\alpha \equiv \frac{1}{\sqrt{\bra{\alpha}{\cal I} - {\cal S}\big({\cal S}^T{\cal S}\big)^{-1}{\cal S}^T\ket{\alpha}}}\,,
\end{equation}
then the QLS follow
\begin{equation}\label{foo13}
\ket{\phi_q} = c_\alpha\left({\cal I} - {\cal S}\big({\cal S}^T\!{\cal S}\big)^{-1}{\cal S}^T\right)\ket{\alpha}\,.
\end{equation}

The normalization factors $c_\alpha$ are closely connected to key observables discussed in previous work. To simplify notations, we denote the projection operator that appears in the definition of $c_\alpha$ in Eq.~(\ref{foo12}) as ${\cal W}\!\equiv\!{\cal I}\!-\!{\cal S}({\cal S}^T{\cal S})^{-1}\!{\cal S}^T\!=\!\sum_\ell\ket{\phi_\ell}\bra{\phi_\ell}$, then we can write
\begin{equation}
c_\alpha = \frac{1}{\sqrt{\bra{\alpha}{\cal W}\ket{\alpha}}}\,,
\end{equation}
i.e.~the normalization factors are the square root of the inverse of the diagonal elements of ${\cal W}$. What is the mechanical interpretation of the operator ${\cal W}$ and of its diagonal elements? The operator ${\cal W}$ was shown in \cite{matthieu_thesis} to play a key role in determining the athermal elastic moduli $C_{ijkl}$ \cite{lutsko} of relaxed Hookean spring networks of unit stiffness, which can be expressed as
\begin{equation}
C_{ijkl} = \Omega^{-1}\bra{\sFrac{\partial r}{\partial \epsilon_{ij}}}{\cal W}\ket{\sFrac{\partial r}{\partial \epsilon_{kl}}}\,,
\end{equation}
with $\Omega$ denoting the system's volume, and $\epsilon$ is the strain tensor. A similar operator to ${\cal W}$ was used in \cite{le_PNAS_2013} in the study of a simple model for supercooled liquids. A dual operator to ${\cal W}$, that projects onto the space of zero modes of ${\cal S}^T{\cal S}$ in floppy materials (i.e.~with $z\!<\! z_c$), was introduced in \cite{ASM_simulations_paper}, and used to construct simulation methods of driven overdamped hard spheres. 

The diagonal elements of ${\cal W}$ can be shown to be equivalent to the `local moduli' recently introduced in \cite{liu_bond_coupling_2017_arXiv} for a single spring in networks of relaxed Hookean springs. In that work the local moduli were proposed as a framework to understand the sensitivity of moduli to removal of springs from simple networks \cite{goodrich_prl_2015}. Following the lines of \cite{liu_bond_coupling_2017_arXiv}, we write the energy $E_\alpha$ associated with imposing a dipolar force on the $\alpha^{\mbox{\tiny th}}$ spring as a sum of squares of the compressions or extensions of all springs, namely
\begin{equation}
E_\alpha = \sFrac{1}{2}\!\sum_\beta\braket{\varphi}{\beta}^2 = \sFrac{1}{2}\bra{\alpha}{\cal S}\big({\cal S}^T{\cal S}\big)^{-1}\!{\cal S}^T\ket{\alpha} = \sFrac{1}{2}\braket{\varphi}{\alpha}\,.
\end{equation}
The fraction of elastic energy stored in all springs except for the $\alpha^{\mbox{\tiny th}}$ spring is
\begin{eqnarray}
\frac{E_\alpha - \sFrac{1}{2}\braket{\varphi}{\alpha}^2}{E_\alpha} & = & \frac{\braket{\varphi}{\alpha} - \braket{\varphi}{\alpha}^2}{\braket{\varphi}{\alpha}} = 1-\braket{\varphi}{\alpha} \nonumber\\
& = & \bra{\alpha}{\cal I} - {\cal S}\big({\cal S}^T{\cal S}\big)^{-1}\!{\cal S}^T\ket{\alpha} \nonumber\\
& = & \bra{\alpha}{\cal W}\ket{\alpha}\,,
\end{eqnarray}
i.e.~it is precisely the $\alpha^{\mbox{\tiny th}}$ diagonal element of ${\cal W}$. The diagonal elements can therefore be understood as indicators of the degree of mechanical coupling of the $\alpha^{\mbox{\tiny th}}$ spring to the rest of the network in which it is embedded. If a certain spring can be pushed against with very little cost of energy in the rest of the system, we deem it weakly coupled. Returning to the discussion about the normalization factors $c_\alpha$, we conclude that large normalization factors  correspond to weakly coupled edges of the network, in the sense described above. We comment further on this point in Sect.~\ref{discussion}. 

In the next Section we describe the result of our numerical simulations, and show that edge-to-edge fluctuations in the values of the normalization factors $c_\alpha$ that distinguish between dipole responses and QLS do not affect the scaling with connectivity of the lengthscale that characterizes both of these objects' spatial decay.



\begin{figure}[!ht]
\centering
\includegraphics[width = 0.50\textwidth]{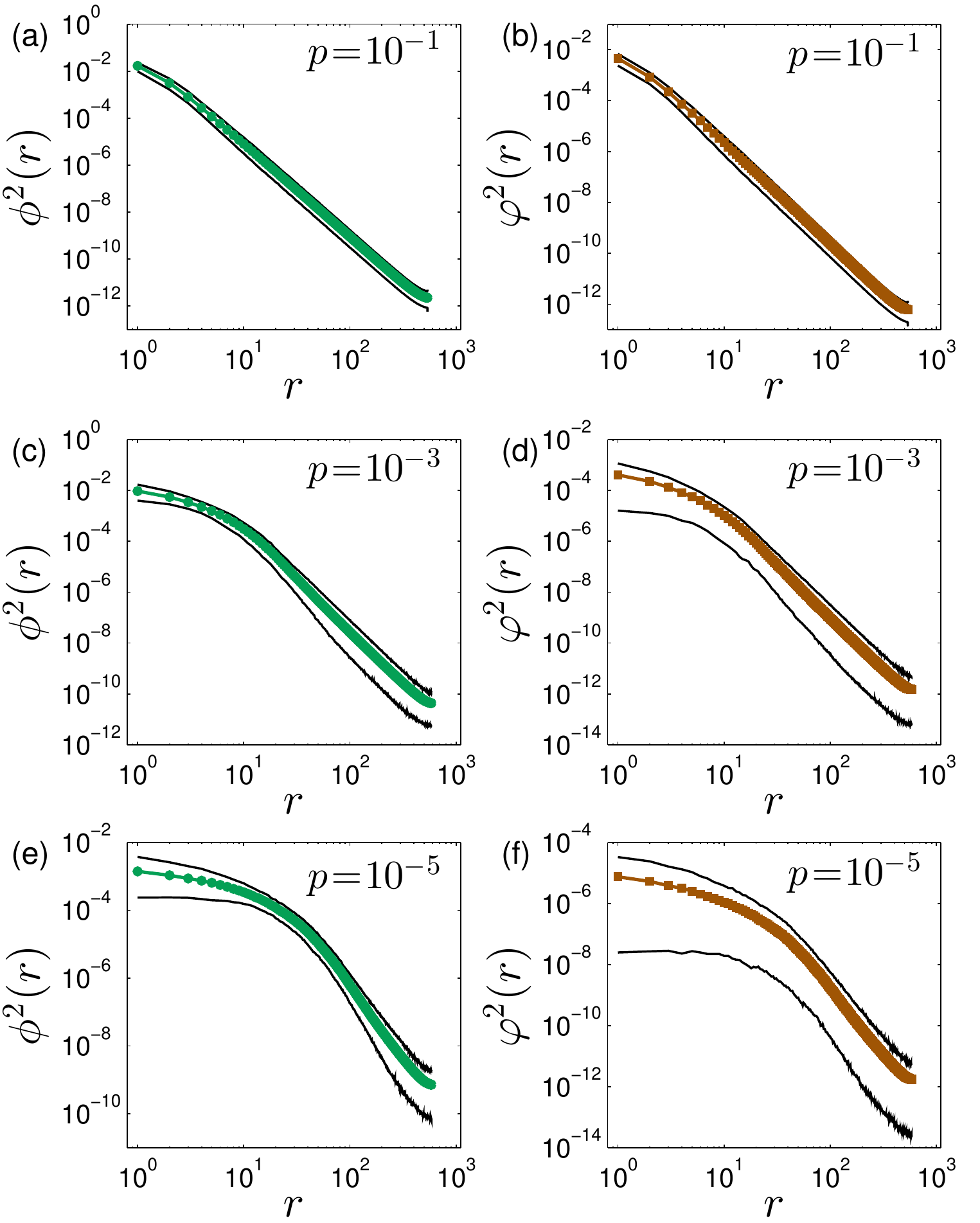}
\caption{\footnotesize Spatial decay of the mean amplitude squared of QLS $\big($filled circles, {\bf (a)},{\bf (c)},{\bf (e)}$\big)$ and of dipole responses $\big($filled squares, {\bf (b)},{\bf (d)},{\bf (f)}$\big)$, plotted against the distance $r$ to the randomly-chosen edge that defines these two objects, see text for further details. The continuous lines enclose the 5$^{\mbox{\tiny th}}$ to 95$^{\mbox{\tiny th}}$ percentiles of the data, and demonstrate that the normalization factors that distinguish between QLS and dipole responses act to substantially reduce the edge-to-edge amplitude fluctuations of QLS.}
\label{quantile_decay_fig}
\end{figure}

\section{Results}
\label{results}

We have calculated the dipole responses $\ket{\varphi}$ and the QLS $\ket{\phi}$ for 600 randomly-selected edges of packing-derived networks generated as explained in Sect.~\ref{methods_section} above. For each randomly selected edge $\ket{\varphi}$ and $\ket{\phi}$ were calculated by solving numerically Eq.~(\ref{foo10}) for $\ket{\varphi}$, and using Eqs.~(\ref{foo12}) and (\ref{foo13}) to obtain the correponding QLS. Notice that here and in the rest of what follows we suppress the subscript `$q$' in the QLS notation.

We first present data that demonstrate how the normalization factors $c_\alpha$ that distinguish between the QLS $\ket{\phi}$ and the dipole responses $\ket{\varphi}$ actually act to substantially decrease edge-to-edge amplitude fluctuations in our ensemble of QLS, compared to the edge-to-edge amplitude fluctuations observed in the dipole responses. We denote by $\phi^2(r)$ and $\varphi^2(r)$ the square of the magnitude of $\ket{\phi}$ and $\ket{\varphi}$, respectively, as a function of the distance $r$ to the edge that defines each of these objects. In Fig.~\ref{quantile_decay_fig} we plot the means of $\phi^2(r)$ (left column, green circles) and $\varphi^2(r)$ (right column, brown squares) vs.~the distance $r$, averaged over our entire calculated ensembles. The pressures from which the networks were derived  (see Sect.~\ref{methods_section} for further details) are indicated by the legends. It is clear that for both objects the crossover to the continuum behavior occurs at a larger lengthscale as $p\!\to\!0$. This length is further discussed below.

We have also outlined in Fig.~\ref{quantile_decay_fig} the areas around the mean spatial decays which cover the 5th-95th percentiles of the data (i.e.~the outlined areas cover 90\% of the data), in order to visualize the reduction of the edge-to-edge amplitude fluctuations of QLS compared to those found for the dipole responses. We find that the relative spread of the dipole responses as represented by our percentile analysis can be larger by a factor of 100 compared to the spread of the QLS, when measured in networks derived from packing at the lowest pressures (compare the outlined areas shown in panels (e) and (f) of Fig.~\ref{quantile_decay_fig}). 

\begin{figure}[!ht]
\centering
\includegraphics[width = 0.45\textwidth]{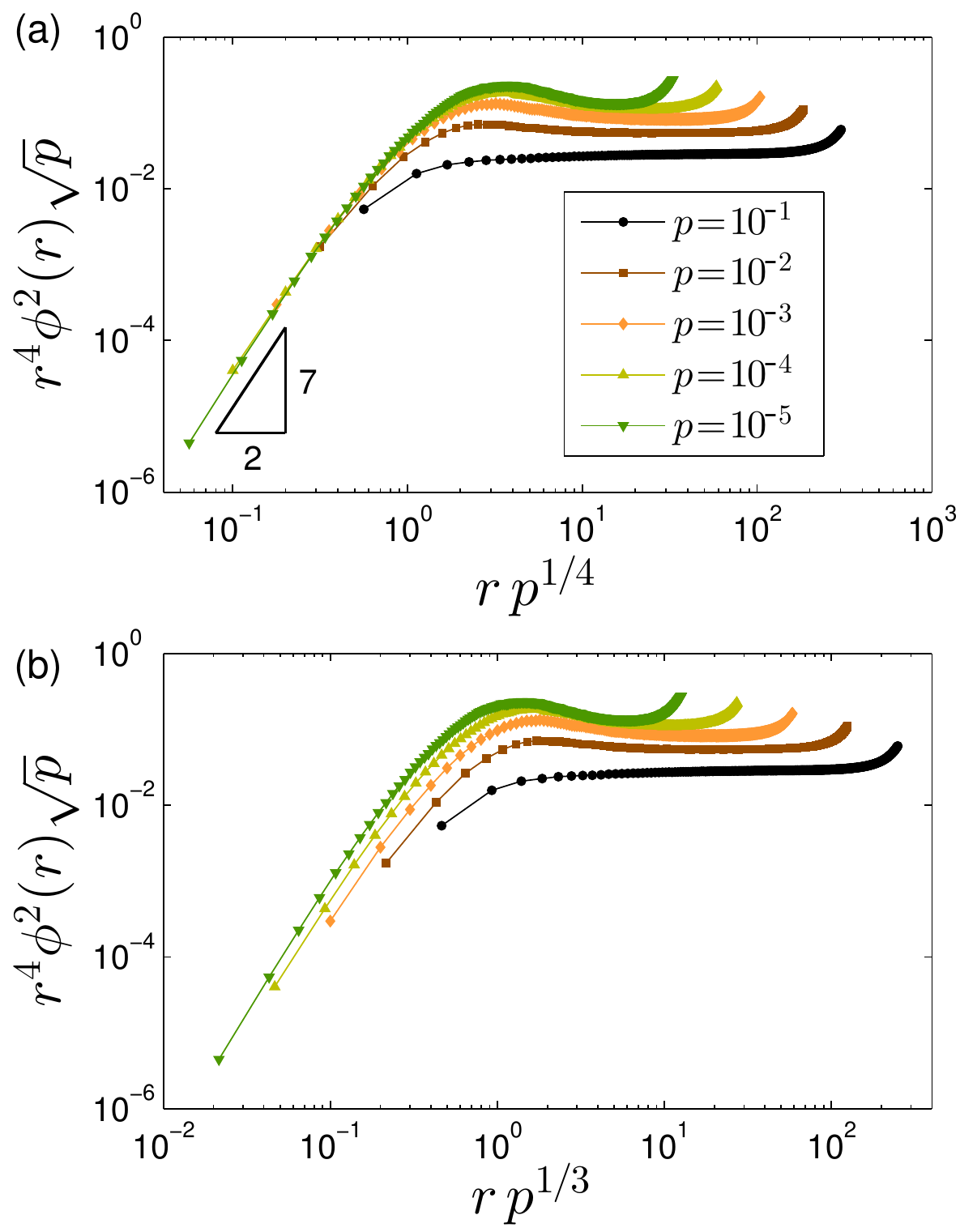}
\caption{\footnotesize {\bf (a)} The products $r^4\phi^2(r)$ rescaled by the characteristic scale of the normalization factors squared $c_\alpha^2\!\sim\! p^{-1/2}$, plotted as a function of the rescaled distances $rp^{1/4}$. {\bf (b)} Same as (a), but plotted as a function of the rescaled length as proposed by \cite{sussman_sss,reply_to_comment}, namely $rp^{1/3}$.}
\label{lengthscale_scaling_fig}
\end{figure}

\begin{figure}[!ht]
\centering
\includegraphics[width = 0.48\textwidth]{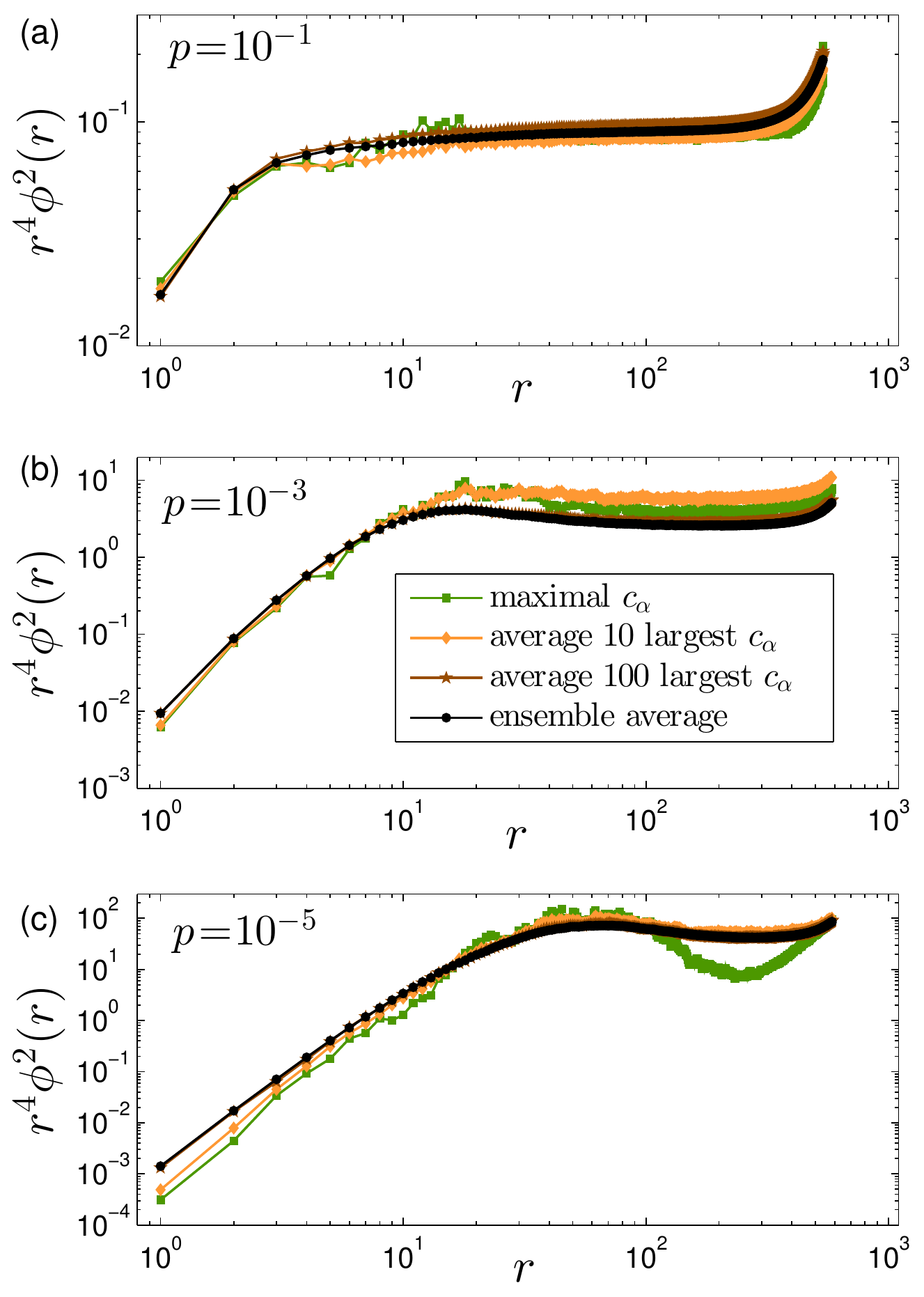}
\caption{\footnotesize Panels {\bf (a)}, {\bf (b)}, and {\bf (c)} show the products $r^4\phi^2(r)$ vs.~distance $r$ measured in networks derived from packings at pressures $p\!=\!10^{-1}$, $p\!=\!10^{-3}$, and $p\!=\!10^{-5}$, respectively. We show the products pertaining to the QLS with the largest normalization factor $c_\alpha$ (green squares), the products averaged over the 10 and 100 QLS with the largest $c_\alpha\!$'s (orange diamonds and brown stars, respectively), and the full average over our entire calculated ensemble of QLS (black circles). }
\label{no_correlation_fig}
\end{figure}

We next focus on resolving the scaling with network connectivity $z$ of the lengthscale that characterizes the spatial decay of  QLS. To this aim, we note first that the amplitude squared of dipole responses $\varphi^2(r)$ was shown in \cite{breakdown} to scale as $r^{-4}$ in the far field (in 2D), with a prefactor that approaches a constant as $z\!\to\! z_c$. This means that in order to achieve a data collapse of the products $r^4\phi^2(r)$ \cite{footnote}, they must be rescaled by the characteristic normalization factors squared $c_\alpha^2$. The latter are estimated as \cite{matthieu_thesis}
\begin{equation}
c_\alpha^2 = \frac{1}{\bra{\alpha}{\cal I}\! -\! {\cal S}\big({\cal S}^T{\cal S}\big)^{-1}\!{\cal S}^T\ket{\alpha}} 
= \frac{1}{\sum_\ell\braket{\phi_\ell}{\alpha}^2} \sim \frac{1}{z - z_c}\,.
\end{equation}
Recalling that in our harmonic-discs-packing-derived networks $z\!-\! z_c\!\sim \sqrt{p}$ \cite{Goodrich_pre_2014}, and assuming that the lengthscale that characterizes the decay of QLS is $\ell_c\!\sim\! p^{-1/4}$, we postulate that $r^4\phi^2(r)\sqrt{p}$ should approach a scaling function ${\cal F}(x)$ if plotted against $rp^{1/4}$, where ${\cal F}(x)\!\sim\! x^\chi$ for small $x$ and ${\cal F}(x)$ approaches a constant for large $x$. In Fig.~\ref{lengthscale_scaling_fig} this hypothesis is tested; we indeed find that as $p\!\to\!0$, $r^4\phi^2(r)\sqrt{p}$ appears to approach a scaling form with $\chi\!\approx\!3.5$, indicating that the lengthscale that characterizes QLS follows $\ell_{\mbox{\tiny SSS}}\!\sim\!\ell_c\!\sim\! (z-z_c)^{-1/2}$. This assertion clearly ignores the uprise in the products $r^4\phi^2(r)$ at large $r$, which is an artifact of the finite size of our systems, and the periodic boundary conditions, as seen in \cite{breakdown}. The identification of $\ell_c$ as the relevant lengthscale is further established in Fig.~\ref{lengthscale_scaling_fig}b, where we test the scaling suggested in \cite{sussman_sss,reply_to_comment} by plotting $r^4\phi^2(r)\sqrt{p}$ vs.~$rp^{1/3}$ to find a clear misalignment of the data.

\begin{figure*}[!ht]
\centering
\includegraphics[width = 0.74\textwidth]{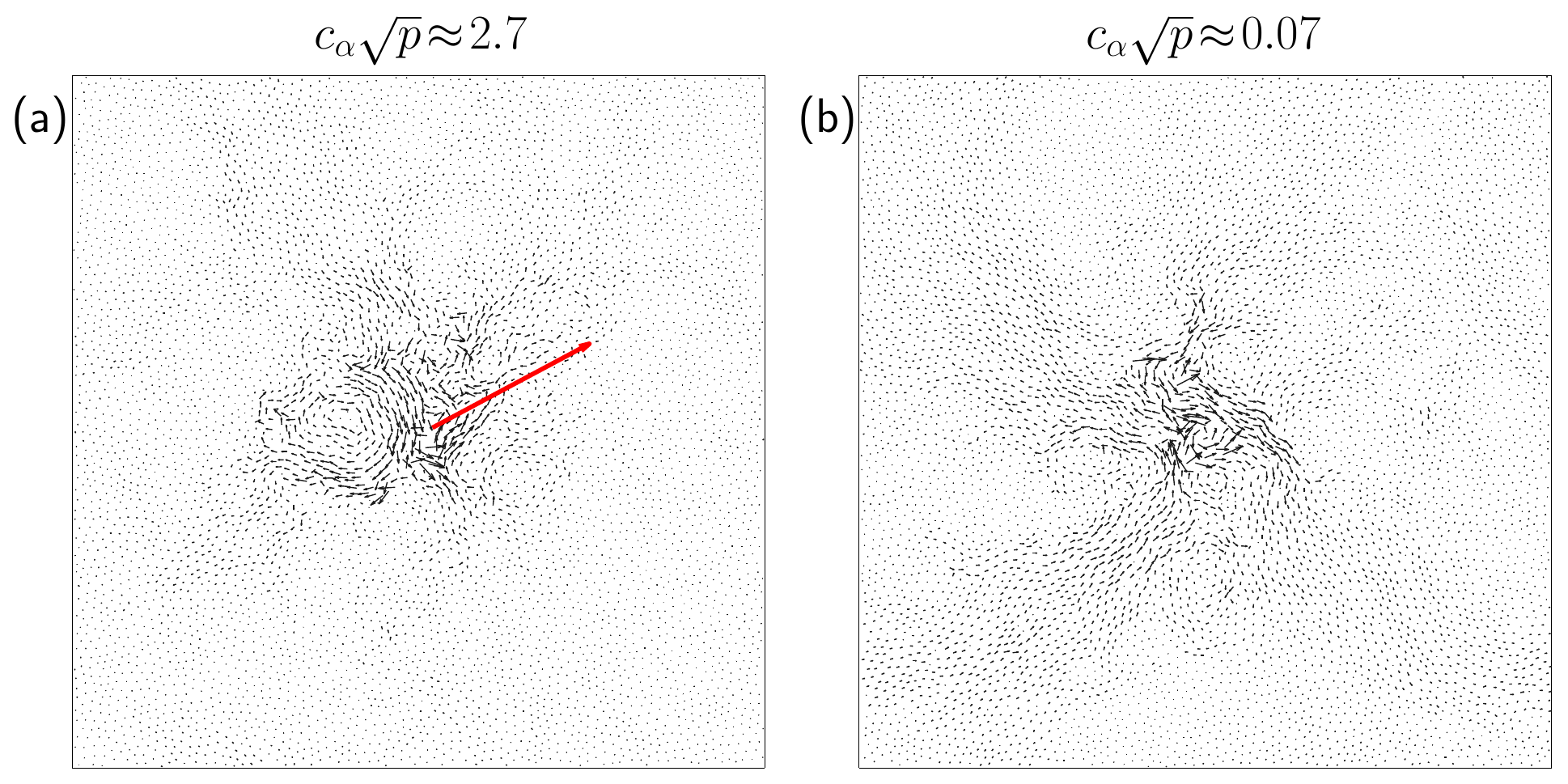}
\caption{\footnotesize Displacement responses $\delta\vec{R}$ as defined by Eq.~(\ref{foo15}), calculated in a random network derived from a packing of $N\!=\!6400$ discs at $p\!=\!10^{-4}\!$, for two different edges with $c_\alpha\sqrt{p}\!\approx\!2.7$ {\bf (a)} and $c_\alpha\sqrt{p}\!\approx\!0.07$ {\bf (b)}. The long red arrow in panel {\bf (a)} represents one of the displacement field components, highlighted and shortened due to its enormous length; in the original, unaltered displacement field its length is 5 times longer. The linear size of the disordered core of both objects is comparable, while their associated normalization factors $c_\alpha$ differ by a factor of almost 40, further supporting that the size of the cores only depends on the network connectivity, and not the normalization factors.}
\label{fields_fig}
\end{figure*}

Up to this point we have established that the normalization factors $c_\alpha$ of QLS (see Eq.~(\ref{foo12}))  lead to a suppression of relative edge-to-edge fluctuations in the amplitude of QLS compared to those seen in amplitudes of dipole responses, and that the scaling with connectivity of $\ell_{\mbox{\tiny SSS}}$ is the same as found for dipole responses in \cite{breakdown}, i.e. $\ell_{\mbox{\tiny SSS}}\!\sim\!\ell_c\!\sim\! (z-z_c)^{-1/2}$. We next check whether there exist correlations between normalization factors and the spatial decay length of their associated QLS. To this aim, we sort the QLS in each ensemble according to their normalization factors $c_\alpha$, and plot in Fig.~\ref{no_correlation_fig} we plot the products $r^4\phi^2(r)$ for the QLS with the largest normalization factors (green squares), and the mean of the same product the 10 and 100 QLS with the largest normalization factors. Each panel displays data calculated in our different ensembles as specified by the values of the pressure reported in the upper right corner. We do not identify a systematic trend that is indicative of correlations in these data; instead, it appears that the length $\ell_{\mbox{\tiny SSS}}$ that characterizes the QLS decay depends only weakly, if at all, on the normalization factors $c_\alpha$.

Further evidence for this apparent independence of $\ell_{\mbox{\tiny SSS}}$ on $c_\alpha$ (for fixed connectivity) can be directly visualized by considering the displacement response to a dipole $\delta\vec{R}$ (see definition in Eq.~(\ref{foo15})) applied to an edge that possesses a large $c_\alpha$, and comparing it to the response pertaining to an edge with a characteristic $c_\alpha$. An example of such a comparison is shown in Fig.~\ref{fields_fig}, where the left (right) panel shows the displacement field pertaining to the large (small) $c_\alpha$. We emphasize that the large-$c_\alpha$ response showed in Fig.~\ref{fields_fig}a is consistently found for other high-$c_\alpha$ edges: it consists of a few (${\cal O}(1)$) very large components near the imposed dipole (shown in red and shortened by a factor of 5 in Fig.~\ref{fields_fig}a), embedded in a background disordered core, whose size depends on the connectivity of the network. In the example of Fig.~\ref{fields_fig} it is also apparent that the disordered core of both displacement responses have comparable sizes, despite that their associated $c_\alpha\!$'s differ by a factor of almost 40, further indicating that $\ell_{\mbox{\tiny SSS}}$ depends on connectivity (as shown above), but not on edge-to-edge fluctuations of $c_\alpha$. 

\vspace{-0.2cm}

\section{Summary and discussion}
\label{discussion}

In this work we have studied in detail the spatial structure of QLS in 2D packing-derived networks. We find strong evidence that the lengthscale $\ell_{\mbox{\tiny SSS}}$ that characterizes the spatial decay of QLS scales with the connectivity difference to the isostatic point as $(z\!-\! z_c)^{-1/2}$, as argued in \cite{comment_liu}, and at odds with the claims made in \cite{sussman_sss,reply_to_comment}. We further showed that the normalization factors $c_\alpha$ that distinguish between QLS and dipole responses substantially suppress edge-to-edge fluctuations of QLS amplitudes, compared to the same fluctuations in dipole responses. We then tested whether averaging the spatial decay of high-$c_\alpha$ QLS leads to observable differences in their decay length, however no systematic effect was~observed. 

We have also showed that a direct visualization of displacement responses to local dipolar forces imposed on high-$c_\alpha$ reveals an interesting pattern: nodes in the immediate vicinity of the dipolar force can have huge displacements compared to their close by neighbors. These large displacements are embedded in a disordered core background whose size appears to be $\ell_{\mbox{\tiny SSS}}$. This finding is reminiscent of the observation of localized excitations in isostatic packings of hard spheres \cite{nonlinear_excitations}, which were shown to be the dominant origin of weak contact forces in such packings. The presence of these weak contacts was later attributed to loosely connected particles in sphere packings, coined ``bucklers" \cite{charbonneau_bucklers}. Interestingly, in a recent work \cite{liu_bond_coupling_2017_arXiv} it was shown that QLS with large normalization factors precisely correspond to edges that connect to buckler particles in the original packing, that are only marginally connected to the rest of the packing. We find consistency with these results when comparing the spatial patterns of displacements that appear upon forcing high-$c_\alpha$ edges.

Our work highlights the importance of considering large systems in studies of diverging lengthscales near unjamming. We find that for networks derived from our two lowest-pressure packings, namely $p\!=\!10^{-4}$ and $p\!=\!10^{-5}$, the distances in connectivity to the Maxwell threshold are on the order of $10^{-2}$. The spatial decay of QLS at these connectivities appear to be close to, but still not converged to, their asymptotic form. Reliably studying lower connectivities would require systems of several millions of particles. 

Finally, the spatial decay profiles we have measured for QLS suggest that for small distances $r\!\lesssim\!\ell_{\mbox{\tiny SSS}}$ from the target edge $\alpha$, the amplitude squared of QLS follows $\phi^2(r)\!\sim\! r^{-1/2}$. This observation is still not understood theoretically, and calls for further numerical tests of its dependence on spatial dimension. 

\vspace{-0.3cm}

\acknowledgements
\vspace{-0.25cm}
We thank D.~Sussmann for discussions. We also thank E.~DeGiuli and G.~D\"uring for useful comments. 


\bibliography{references_lerner_jamming}

\begin{thebibliography}{45}
\bibitem [{foo()}]{footnote}%
  \BibitemOpen
  \href@noop {} {\bibinfo  {journal} {The strong spatial decay of QLSsSS
  results in a variation of their amplitude squared over up to 10 orders of
  magnitude, as seen in Fig.~\ref{quantile_decay_fig}. For this reason, the
  identification of the fine details of the spatial structure of QLSsSS is best
  achieved by considering the products $r^4\phi^2(r)$}\ }\BibitemShut {NoStop}%
\end{thebibliography}

\end{document}